\documentstyle[12pt]{article}
 \advance\oddsidemargin by -1in
 \advance\evensidemargin
 by -1in
 \marginparsep
 0.4cm
 \advance\topmargin by
 -0.0in
 \makeindex

 \newcommand\la{\langle}
 \newcommand\ra{\rangle}
 \newcommand\beq{\begin{equation}}
 
 \newcommand\eeq{\end{equation}}
 \newcommand\beqn{\begin{eqnarray}}
 \newcommand\eeqn{\end{eqnarray}}
 \newcommand{\doublespace} {
 \renewcommand{\baselinestretch} {1.6}
 \large\normalsize}

\begin{document}

%\vspace*{5cm}

\title{\bf Nuclear Shadowing and Coherence
Length for Longitudinal and Transverse Photons}

%\maketitle

%\centerline{\Large \bf Nuclear Shadowing and Coherence
%Length} 

%\medskip

%\centerline{\Large \bf for Longitudinal and Transverse Photons}

%\vspace{.5cm}
%\author

\vspace{1.3cm}

\begin{center}

\author{\large Boris~Kopeliovich$^{1,3}$, J\"org Raufeisen$^{1,2}$,
and Alexander Tarasov$^{1,3}$}

%\date{\empty}

\maketitle

 $^1${\sl Max-Planck Institut f\"ur Kernphysik, Postfach
103980, 69029 Heidelberg}\\[0.1cm]
$^2${\sl Institut f\"ur Theor. Physik der Universit\"at ,
Philosophenweg 19, 69120 Heidelberg}\\[0.1cm]
$^3${\sl Joint Institute for Nuclear Research, Dubna,
141980 Moscow Region, Russia}

\end{center}

\vspace{1cm}

\begin{abstract}
Motivated by the recent results
for DIS off nuclei from the HERMES experiment
we have performed a systematic
study of shadowing for transverse and longitudinal 
photons. We found that the coherence length which
controls the onset of nuclear shadowing at small 
$x_{Bj}$ is much longer for longitudinal than transverse 
photons, and is much shorter for shadowing of gluons. 
The light-cone Green function approach we apply 
properly treats shadowing in the transition region 
$x_{Bj}>0.01$. It also 
incorporates the nonperturbative 
effects and is legitimate at small $Q^2$. 
We calculate nuclear shadowing
and compare with data from the HERMES and NMC experiments.
Although we expect different nuclear shadowing for longitudinal 
and transverse photons,  numerically it cannot explain the
strong effect observed by the HERMES collaboration.
\end{abstract}

\doublespace

\newpage

\section{Introduction}\label{intro}

Recently the collaboration HERMES released data 
\cite{hermes} for 
shadowing in inclusive positron scattering off nuclei
at medium high energies and $Q^2$. The results expose 
few unusual features. The cross sections on nuclear targets,
nitrogen and helium-3, at small $x_{Bj}\approx 0.02$
and $Q^2<1\,GeV^2$ was found to be 
substantially more shadowed that one could expect extrapolating
available data at higher $Q^2$ and energies.
Unusual is also the observed enhancement of shadowing with $Q^2$.
Interpreted in terms of different shadowing for transverse and
longitudinal photons it was concluded in \cite{hermes} 
that $\sigma_L$ is
enhanced, while $\sigma_T$ is suppressed on nitrogen by at least
factor of two compared to deuteron target.

These data drew attention to the fact that very few data are
available in this kinematical region. Moreover, no reliable
theoretical calculations are done yet. The approach based
on the nonlinear evolution equations \cite{glr,mq} needs knowledge
of the nuclear parton distribution at a medium-hard scale
which is to be guessed, and is anyway outside the kinematical range
we are interested in. A more promising is the 
intuitive approach treating nuclear 
effects in the spirit of vector dominance model (VDM) \cite{bauer} 
as shadowing for the total cross section of hadronic fluctuations 
of the virtual photon (see {\it e.g.} in \cite{fs}).
However, the perturbative QCD treatment of the photon fluctuation
can be applied only at high $Q^2$, while VDM is sensible only
at small $Q^2\to 0$. 

A progress was done recently \cite{kst2} on extension
of perturbative QCD methods to the region of small $Q^2$
where the quarks in photon fluctuations cannot be treated as free.
The nonperturbative inter-quark interaction was 
explicitly introduced and new light-cone $\bar qq$ wave functions
were derived which recover the well know perturbative ones at 
large $Q^2$.

Nuclear shadowing is controlled by the interplay between
two fundamental quantities.

\begin{itemize}
\item{The lifetime of photon fluctuations, or coherence time.
Namely, shadowing is possible only if the coherence 
time exceeds the mean internucleon spacing in nuclei,
and shadowing saturates (for a given Fock component)
if the coherence time substantially exceeds the nuclear 
radius.}
\item{Equally important for shadowing is the transverse
separation of the $\bar qq$. In order to be shadowed the 
$\bar qq$-fluctuation of the photon has to interact with
a large cross section. 
As a result of colour transparency \cite{zkl,bm,bbgg}, small size
dipoles interact only weakly and are therefore less shadowed. The dominant
contribution to shadowing comes from  the large
aligned jet configurations
\cite{ajm,fs} of the pair.}
\end{itemize}
The mean lifetime of a $\bar qq$ fluctuation in vacuum
calculated in section~\ref{vacuum} turns out to be zero for transverse
photons. This strange result is a consequence of an incorrect
definition. 

In sections~\ref{lc-pqcd} and \ref{lc-npr} we 
propose a more sophisticated treatment of the coherence length 
or the fluctuation lifetime relevant for shadowing.  
The mean coherence time for the $\bar qq$ Fock 
state is evaluated using the perturbative and nonperturbative 
wave functions. The salient observation is that 
the coherence length is nearly three times longer 
for longitudinal than for transverse photons. At the same time, both
are substantially different from the usual prescription 
$l_c=(2m_Nx_{Bj})^{-1}$. The coherence length is found to 
vary steeply with $Q^2$ at fixed $x_{Bj}$ and small $Q^2$.

The coherence time for a $|\bar qqG\ra$ Fock component controlling
nuclear shadowing for gluons is calculated in section~\ref{lc-g}.
It turns out to be much shorter than for $|\bar qq\ra$ components,
therefore, onset of gluon shadowing is expected at smaller $x_{Bj}$
than for quarks.

The transition region between no-shadowing at $x_{Bj}\sim 0.1$
and saturated (for the $|\bar qq\ra$ component) shadowing
at very small $x_{Bj}$ is most difficult for theory. 
The impact parameter representation assigns definite cross
sections to the fluctuations, but no definite mass
which one needs to calculate the phase shift. On the other
hand, the eigenstates of the mass matrix cannot be
associated with any definite cross section.
This controversy was settled 
within the light-cone Green function approach \cite{krt,rtv,zakh}.
In section~\ref{shad} we rely on
this approach to calculate nuclear 
shadowing in the kinematical region of the HERMES experiment.
The nonperturbative light-cone
wave functions and the realistic phenomenological dipole
cross section are important at low $Q^2$.
We are unable to reproduce the 
data from the HERMES experiment, although our 
parameter-free calculations are
in a good agreement with NMC data.  We do not expect any dramatic
enhancement of the longitudinal cross section compared to 
the transverse one. At the same time, the calculated transverse 
cross section for nitrogen is shadowed by about $20\%$ only, 
much less than the HERMES data need.

\section{The mean coherence length}\label{lc}

\subsection{The lifetime for a perturbative \boldmath$\bar qq$
fluctuation in vacuum}\label{vacuum}

A photon of virtuality $Q^2$ and energy $\nu$
can develop a hadronic fluctuation for a lifetime,
\beq
l_c=\frac{2\,\nu}{Q^2+M^2}= 
\frac{P}{x_{Bj}\,m_N} =
P\,l_c^{max}\ ,
\label{1.1}
\eeq
where Bjorken $x_{Bj}=Q^2/2m_N\nu$,
$M$ is the effective mass of the fluctuation,
factor $P=(1+M^2/Q^2)^{-1}$, and $l_c^{max}=1/m_Nx_{Bj}$. 
The usual approximation is to assume that
$M^2 \approx Q^2$ since $Q^2$ is the only large 
dimensional scale available. In this case $P=1/2$.

The effective mass of a noninteracting $q$ and $\bar q$ 
is well defined, $M^2=(m_q^2+k_T^2)/\alpha(1-\alpha)$,
where $m_q$ and $k_T$ and $\alpha$ are the mass, transverse
momentum and fraction of the
light-cone momentum of the photon carried by the quark.
Therefore, $P$ has a simple form,
\beq
P(k_T,\alpha)=\frac{Q^2\,\alpha\,
(1-\alpha)}{k_T^2+\epsilon^2}\ ,
\label{1.2}
\eeq
where
\beq
\epsilon^2 = \alpha(1-\alpha)Q^2 + m_q^2\ .
\label{1.3}
\eeq
To find the mean value of the fluctuation lifetime in vacuum 
one should average (\ref{1.2}) over $k_T$ and $\alpha$ 
weighted with the wave function squared of
the fluctuation,
\beq
\la P\ra_{vac}=
\frac{\Bigl\la \Psi^{\gamma^*}_{\bar qq}
\Bigl|P(k_T,\alpha)\Bigr|
\Psi^{\gamma^*}_{\bar qq}\Bigr\ra}
{\Bigl\la \Psi^{\gamma^*}_{\bar qq}\Bigl|
\Psi^{\gamma^*}_{\bar qq}\Bigr\ra}\ .
\label{1.3a}
\eeq

The perturbative distribution function for the $\bar qq$
has the form \cite{ks,bks,nz},
\beq
\Psi^{T,L}_{\bar qq}(\vec r_T,\alpha)=
\frac{\sqrt{\alpha_{em}}}{2\,\pi}\,
\bar\chi\,\widehat O^{T,L}\,\chi\,K_0(\epsilon r_T)
\label{1.4}
\eeq
Here $\chi$ and $\bar\chi$ are the spinors of the quark
and antiquark respectively.
$K_0(\epsilon r_T)$ is the modified Bessel function.
The operators $\widehat O^{T,L}$ have the form,
\beq
\widehat O^{T}=m_q\,\vec\sigma\cdot\vec e +
i(1-2\alpha)\,(\vec\sigma\cdot\vec n)\,
(\vec {e}\cdot \vec\nabla_{T})
+ (\vec\sigma\times\vec e)\cdot\vec\nabla_{T}\ ,
\label{1.5}
\eeq
\beq
\widehat O^{L}= 2\,Q\,\alpha(1-\alpha)\,\vec\sigma\cdot\vec n\ ,
\label{1.6}
\eeq
where the two-dimensional operator $\vec\nabla_{T}$
acts on the transverse coordinate $\vec r_T$;
$\vec n=\vec p/p$ is a unit vector parallel to
the photon momentum; $\vec e$ is the polarization vector
of the photon.

The normalization integral in the denominator in the
{\it r.h.s.} of (\ref{1.3a})
diverges at large $k_T$ for transversely polarized photons, 
therefore we arrive at the unexpected result
$\la P^T\ra_{vac}=0$.

\subsection{Coherence length in nuclear medium}\label{lc-pqcd}

This puzzling conclusion can be interpreted as a result
overwhelming the fluctuations of a transverse photon by
heavy $\bar qq$ pairs with very large $k_T$. Such heavy fluctuations
indeed have a very short lifetime. However, they also have a vanishing
transverse size $r_T\sim 1/k_T$ and interaction cross section.
Therefore, such fluctuation cannot be resolved by the interaction and do
not contribute to the DIS cross section.
To get a sensible result one should properly define the averaging
procedure. We are interested in the fluctuations which contribute
to nuclear shadowing, {\it i.e.} they have to interact at least twice.
Correspondingly, the averaging procedure 
has to be redefined as,
\beq
\la P\ra_{shad}=
\frac{\Bigl\la f(\gamma^*\to\bar qq)
\Bigl|P(k_T,\alpha)\Bigr|
f(\gamma^*\to\bar qq)\Bigr\ra}
{\Bigl\la f(\gamma^*\to\bar qq)\Bigl|
f(\gamma^*\to\bar qq)\Bigr\ra}\ ,
\label{1.6a}
\eeq
where $f(\gamma^*\to\bar qq)$ is the amplitude of diffractive
dissociation of the virtual photon on a nucleon 
$\gamma^*\,N\to\bar qq\,N$.

Thus, one should include in the weight the 
interaction cross section squared $\sigma^2_{\bar qq}(r_T,s)$, 
where $r_T$ is the transverse separation and $s=2m_N\nu-Q^2+m_N^2$.
Then, the mean value of factor $P(\alpha,k_T)$ reads,
\beq
\left\la P^{T,L}\right\ra=\frac{\int_0^1d\alpha\int d^2r_1d^2r_2
\left[\Psi_{q\bar q}^{T,L}\left(\vec r_2,\alpha\right)\right]^*
\sigma^N_{q\bar q}\left(r_2,s\right)
\widetilde P\left(\vec r_2-\vec r_1,\alpha\right)
\Psi_{q\bar q}^{T,L}\left(\vec r_1,\alpha\right)
\sigma^N_{q\bar q}\left(r_1,s\right)}
{\int_0^1d\alpha\int d^2r\,
\left|\Psi_{q\bar q}^{T,L}\left(\vec r_,\alpha\right)
\sigma^N_{q\bar q}\left(r,s\right)\right|^2}
\label{1.7}
\eeq
with
\beq
\widetilde P\left(\vec r_2-\vec r_1,\alpha\right)=
\int\frac{d^2k_T}{\left(2\pi\right)^2}\,
{\exp\left(-i\,\vec k_T\cdot\left( \vec r_2-\vec r_1\right)\right)}
{P\left(\alpha,k_T\right)}.
\label{1.8}
\eeq

Using expression (\ref{1.2}) one obtains for
a non interacting  $q\bar q$-pair,
\beq
\widetilde P\left(\vec r_2-\vec r_1,\alpha\right)=
\frac{Q^2\alpha\left(1-\alpha\right)}
{2\pi}K_0\left(\varepsilon\left|\vec r_2
-\vec r_1\right|\right).
\label{1.9a}
\eeq

As a simple estimate for the mean value 
(\ref{1.7}) one can use the small-$r_T$
approximation for the dipole cross section 
$\sigma_{\bar qq}(r_T,s)=C(s)\,r_T^2$. 
The factor $C(s)$ does not enter the
result since it cancels in (\ref{1.7}).
We obtain for transverse and longitudinal photons,
\beq
\left\la P^T\right\ra =
\frac{2\,Q^2}{3}\,\,
\frac{\int\limits_{0}^{1}
d\,\alpha\,
(1-\alpha)\,\alpha\,
\Bigl(
\Bigl[\alpha^2+(1-\alpha)^2\Bigr]\,
\Bigl/\,\epsilon^6
+\frac{7}{8}\,m_q^2(1-\alpha)\,\alpha\,\Bigl/\,\epsilon^8
\Bigr)
}
{\int\limits_{0}^{1}
d\,\alpha\,
\Bigl(
\Bigl[\alpha^2+(1-\alpha)^2\Bigr]\,
\Bigl/\,\epsilon^4\
+\frac{2}{3}\,m_q^2\Bigl/\,\epsilon^6
\Bigr)}
 ;
\label{1.10}
\eeq
\beq
\left\la P^L\right\ra =
\frac{7\,Q^2}{8}\,\,
\frac{\int\limits_{0}^{1}
d\,\alpha\,
(1-\alpha)^3\,\alpha^3\,
\Bigl/\,\epsilon^8}{\int\limits_{0}^{1}
d\,\alpha\,
(1-\alpha)^2\,\alpha^2\,
\Bigl/\,\epsilon^6}\ ,
\label{1.10a}
\eeq
respectively.

\begin{figure}[tbh]
\includegraphics{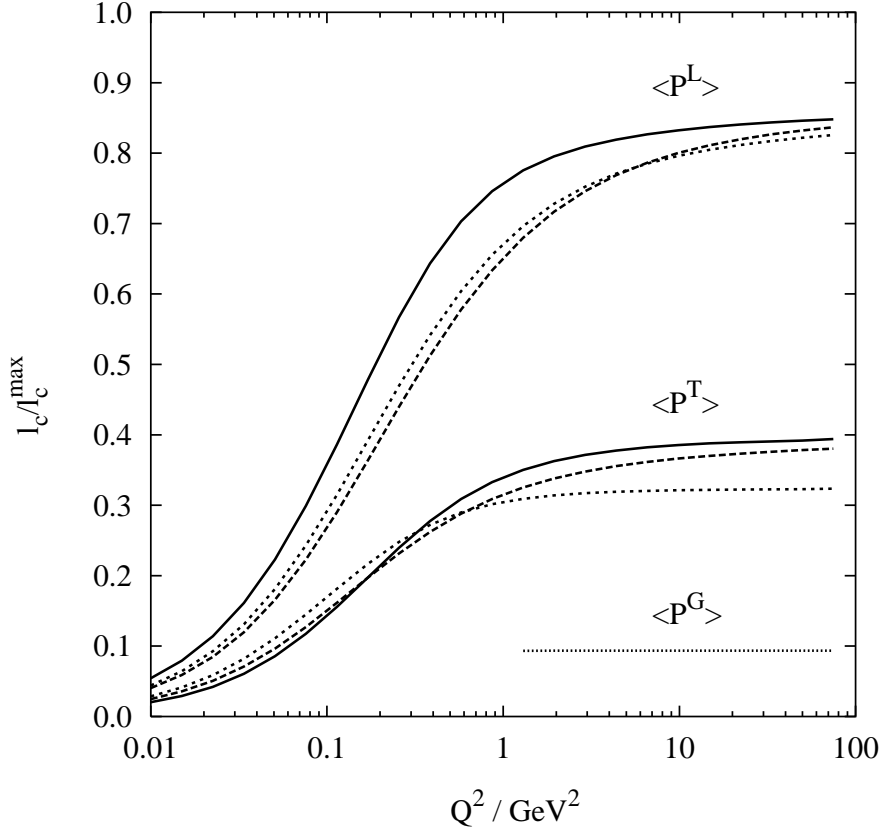}
\begin{center}
\vspace{11.3cm}
\parbox{13cm}
{\caption[shad1]
{\sl $Q^2$ dependence of the factor $\left\la P\right\ra=
l_c/l_c^{max}$ defined in (\ref{1.1}) at $x_{Bj}=0.01$. 
for $\bar qq$ fluctuations of transverse 
and longitudinal
photons, and for $\bar qqG$ fluctuation, from the top to 
bottom, respectively. Dotted curves correspond to
calculations with perturbative wave functions and an approximate dipole
cross section $\propto r_T^2$. Dashed curves are the same, except
the realistic parameterization (\ref{1.9}).
The solid curves show the most realistic case 
based on the nonperturvative wave function (\ref{2.2}).
The coherence length for gluons calculated in sec. (\ref{lc-g}) 
is also shown.}
\label{l-Q2}}
\end{center}
\end{figure}

We calculated the factor $\la P^{T,L}\ra$ as function of $Q^2$ 
at $x_{Bj}=0.01$ which is close to the minimal value in the HERMES data.
We used the effective quark mass $m_q=0.2\,GeV$ which corresponds
to the confinement radius and allows to reproduce data on nuclear 
shadowing \cite{krt}.
Our results depicted in Fig.~\ref{l-Q2} by dotted lines are 
quite different from the naive estimate $P^{T,L}=1/2$. 
Besides, $P^L$ turns out to be substantially longer than $P^T$.
This indicates that a longitudinally polarized photon develops
lighter fluctuations than a transverse one. Indeed, the effective mass
$M$ is maximal for asymmetric pairs, {\it i.e.} when
$\alpha$ or $1-\alpha$ are small. However, such fluctuations
are suppressed in longitudinal photons by the
wave function (\ref{1.6}).

\begin{figure}[tbh]
\includegraphics{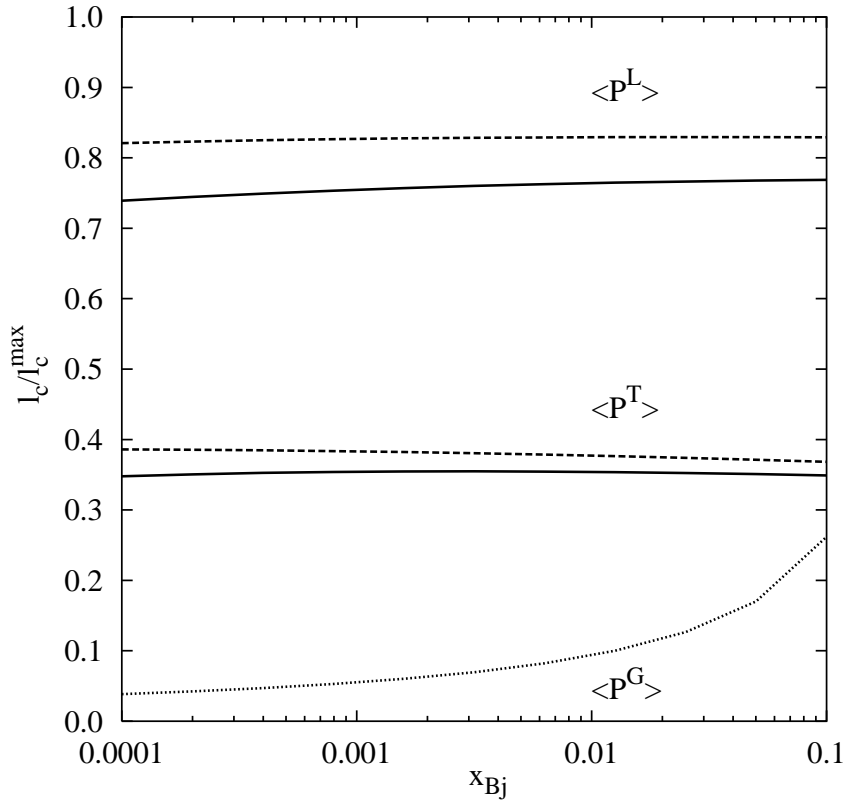}
\begin{center}
\vspace{10.5cm}
\parbox{13cm}
{\caption[shad1]
{\sl $x_{Bj}$ dependence of the factor $\left\la P^{T,L}\right\ra$
and $\left\la P^{G}\right\ra$ 
defined in (\ref{1.1}) corresponding to the coherence length 
for shadowing of transverse and longitudinal
photons and gluon shadowing, respectively. Solid and dashed curves
correspond to $Q^2=4$ and $40\,GeV^2$.}
\label{l-x}}
\end{center}
\end{figure}

The dependence of $\left\la P^{T,L}\right\ra$ on $x_{Bj}$ 
depicted in Fig.~\ref{l-x} for $Q^2=4$ and $40\,GeV$
is rather smooth. Therefore, the coherence length 
varies approximately as $l_c\propto 1/x_{Bj}$.

The simple approximation $\sigma_{\bar qq} \propto r_T^2$ is not
realistic since nonperturbative effects affect the large-$r_T$ behaviour.
Motivated by the phenomenon of confinement one should expect 
that gluons cannot propagate far away and 
the cross section should level off at large $r_T$.
We use the modification \cite{kst2}
of the energy dependent phenomenological dipole cross section
$\sigma_{\bar qq}^N(r_T,s)$ suggested in \cite{gw},
\beq
\sigma^N_{q\bar{q}}\left(r_T,s\right)
=\sigma_0(s)
\left[1-{\rm exp}\left(-\frac{r_T^2}
{r_0^2\left(s\right)}\right)\right]
\label{1.9}
\eeq
where $r_0\left(s\right)=0.88\,(s/s_0)^{-0.14}\,fm$,
$s_0=1000\,GeV^2$.
This cross section is proportional to $r_T^2$ at small
$r_T\to 0$, but is constant at
large $r_T$. The energy dependence correlates with $r_T$,
at small $r_T$ the dipole cross section rises steeper 
with energy than at large separations:
\beq
\sigma_0(s)=\sigma_{tot}^{\pi p}(s)\left(1+
\frac{3\,r_0^2\left(s\right)}{8\left<r^2_{ch}\right>_{\pi}}\right)
\quad,\quad\left<r^2_{ch}\right>_{\pi}=0.44\,{\rm fm}^2,
\label{1.10c}
\eeq
where
\beq
\sigma_{tot}^{\pi p}(s)=23.6\,(s/s_0)^{0.08}\,{\rm mb}.
\label{1.11}
\eeq
With this choice of $\sigma_{tot}^{\pi p}(s)$ one
automatically reproduces the total cross section for pion proton 
scattering, while the parameterization from \cite{gw} cannot be applied to
hadronic cross sections.
Thus, cross section (\ref{1.9}) is better designed for
low and medium large $Q^2< 10-20\,GeV^2$, while at high $Q^2$
the parameterization \cite{gw} works better.

Eq.~(\ref{1.7}) can be represented in the form,
\beq
\Bigl\la P^{T,L}\Bigr\ra =
\frac{N^{T,L}}{D^{T,L}}\ ,
\label{1.12}
\eeq
The angular integrations in (1.7) for the denominators 
$D^{T,L}$ are trivial and for the
numerators $N^{T,L}$ one
uses the relation \cite{gr},
\beq
K_0\left(\varepsilon\left|\vec r_1
-\vec r_2\right|\right)=K_0\left(\varepsilon r_>\right)I_0\left(\varepsilon
r_<\right)+2\sum_{m=1}^\infty e^{im\phi}
K_m\left(\varepsilon r_>\right)I_m\left(\varepsilon
r_<\right),
\label{1.12a}
\eeq
where $r_>=\max\left(r_1,r_2\right)$, $r_<=\min\left(r_1,r_2\right)$,
$\cos\phi=\vec r_1\cdot\vec r_2/(r_1r_2)$ and $I_m(z)$ are the
modified Bessel functions of first kind 
(Bessel function of imaginary variable). 
It is clear from this relation
 that after angular integration only one term in the sum gives a
non-vanishing contribution. We finally obtain for transverse photons,
\beqn\nonumber
N_p^{T}&=&2\,Q^2\int\limits_0^1
d\alpha\,\alpha\,\left(1-\alpha\right)
\int\limits_0^\infty dr_2\ r_2\int\limits_0^{r_2}
dr_1\ r_1\left\{m_q^2\,
K_0^2\left(\varepsilon r_2\right)\,K_0\left(\varepsilon r_1\right)
\,I_0\left(\varepsilon r_1\right)\right.\\
&+&\left.\left[\alpha^2+
\left(1-\alpha\right)^2\right]\,\varepsilon^2\,
K_1^2\left(\varepsilon r_2\right)\,
K_1\left(\varepsilon r_1\right)
\,I_1\left(\varepsilon r_1\right)\right\}
\,\sigma_{q\bar q}^N\left(s,r_1\right)\,
\sigma_{q\bar q}^N\left(s,r_2\right)\ ,
\label{1.13}
\eeqn
\beq
D_p^{T}=\int\limits_0^1 d\alpha
\int\limits_0^\infty dr\ r\left\{
m_q^2\,
K_0^2\left(\varepsilon r\right)
+\left[\alpha^2+\left(1-\alpha\right)^2\right]\varepsilon^2
\,K_1^2\left(\varepsilon r\right)\right\}
\Bigl[\sigma_{q\bar q}^N\left(s,r\right)\Bigr]^2\ ,
\label{1.14}
\eeq
and for longitudinal photons,
\beq
N_p^{L}=2\,Q^2\int d\alpha\,\alpha^3\,\left(1-\alpha\right)^3
\int\limits_0^\infty dr_2r_2\int\limits_0^{r_2} dr_1r_1
K_0^2\left(\varepsilon r_2\right)K_0\left(\varepsilon r_1\right)
I_0\left(\varepsilon r_1\right)
\sigma_{q\bar q}^N\left(s,r_1\right)\sigma_{q\bar q}^N\left(s,r_2\right)\ ,
\label{1.15}
\eeq
\beq
D_p^{L}=\int d\alpha
\int\limits_0^\infty dr r
\alpha^2\left(1-\alpha\right)^2
K_0^2\left(\varepsilon r\right)
\sigma_{q\bar q}^N\left(s,r\right)^2\ .
\label{1.16}
\eeq

The factor $\Bigl\la P^{T,L}(x,Q^2)\Bigr\ra$ calculated in this way 
is depicted by dashed lines in Fig.~\ref{l-Q2} as
function of $Q^2$ at $x_{Bj}=0.01$. 
It is not much different from the previous simplified
estimate demonstrating low sensitivity to the form
of the dipole cross section.

\begin{figure}[tbh]
\includegraphics{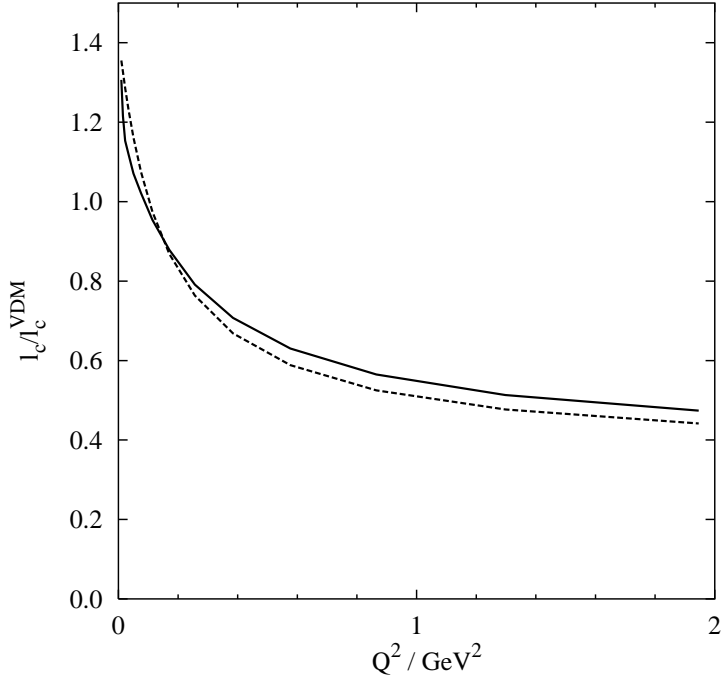}
\begin{center} 
\vspace{9.5cm}
\parbox{13cm}
{\caption[shad1]
{\sl $Q^2$ dependence of ratio of $\la l_c^T\ra$  calculated
with Eq.~(\ref{1.7}) and $m_q=200\,MeV$ to
$l_c^{VDM}$ calculated with Eq.~(\ref{1.1}) and $M=m_{\rho}$.}
\label{vdm}}
\end{center}
\end{figure}

It is instructive to compare our predictions with the VDM 
which is usually supposed to dominate
at small $Q^2 \leq m_{\rho}^2$. The corresponding coherence length
$l_c^{VDM}$ is given by (\ref{1.1}) with $M=m_{\rho}$.
The ratio of $l_c^T$ calculated with the nonperturbative wave function
to $l_c^{VDM}$ as function of $Q^2$ is shown by solid curve in Fig.~\ref{vdm}.
It demonstrates an unexpectedly precocious
violation of VDM at quite low $Q^2$. We also calculated $l_c^T$ with
the perturbative wave function, but with a massive quark. With $m_q=200\,MeV$
it mimics the nonperturbative effects quite well, as one can see from 
Fig.~\ref{vdm}. 

\subsection{Propagation of interacting \boldmath$\bar qq$}
\label{lc-npr}

Although the quarks should be treated perturbatively
as nearly massless, at the endpoints
$\alpha$ or $1-\alpha \to 0$ the mean $\bar qq$ transverse separation
$r_T\sim 1/\epsilon$ 
becomes huge $\sim 1/m_q$. This contradicts the concept of confinement
and should be regularized either by explicit introduction of a nonperturbative 
interaction between $q$ and $\bar q$ \cite{kst2}, or 
as is done below, introducing an effective quark mass.

Apparently, it is not legitimate to use the perturbative 
$\bar qq$ wave functions (\ref{1.4}) -- (\ref{1.5})
at low $Q^2$ where the nonperturbative
interaction between $q$ and $\bar q$ becomes important.
This interaction squeezes the $\bar qq$ wave packet,
{\it i.e.} increases the intrinsic 
transverse momentum and the effective
mass of the pair. By 
contributing to the effective mass of the $\bar qq$,
the nonperturbative interaction
breaks down the validity of the kinematical expression for $M$ or
Eq.~(\ref{1.2}). Even at high $Q^2$ at the endpoints
$\alpha$ or $1-\alpha \to 0$ the mean $\bar qq$ transverse separation
$r_T\sim 1/\epsilon$ becomes huge $\sim 1/m_q$ and nonperturbative
corrections may be important.
One can try to mimic these effects by an effective quark
mass, as is done above, but one never knows how good this
recipe is. 
To take the nonperturbative effects into account we
use the light-cone Green function formalism suggested in
\cite{kst2} generalizing the perturbative description \cite{krt}
of nuclear shadowing in DIS.

Propagation of an interacting $\bar qq$ pair in vacuum
with initial separation $\vec r_1$ at the point with 
longitudinal coordinate $z_1$ up to point $z_2$ 
with final separation $\vec r_2$
is described by a 
Green function $G_{\bar qq}(\vec r_2,z_2;\vec r_1,z_1)$
which is a solution of a 
two dimensional Schr\"odinger equation,
\beq
i\frac{d}{dz_2}\,G^{vac}_{\bar qq}(z_1,\vec r_1;z_2,\vec r_2)\,=\,
\frac{\epsilon^2 - \Delta_{r_2}
+ a^4(\alpha)\,r_2^2}{2\,\nu\,\alpha\,(1-\alpha)}\ 
G^{vac}_{\bar qq}(z_1,\vec r_1;z_2,\vec r_2)\ .
\label{2.1}
\eeq
The nonperturbative oscillator potential contains function
$a(\alpha) = a_0+a_1\,\alpha\,(1-\alpha)$ with parameters $a_0$
and $a_1$ fixed by data \cite{kst2}. Data on total photoabsorption
cross section, diffraction and nuclear shadowing are well
described with,
\beq
a^2\left(\alpha\right)=v^{1.15}(112MeV)^2
+\left(1-v\right)^{1.15}
(165MeV)^2\alpha\left(1-\alpha\right)\ ,
\label{2.1a}
\eeq
where $v$ is any number satisfying $0 < v < 1$.

The Green function allows to calculate the nonperturbative
wave function for a $\bar qq$ fluctuation,
\beq
\Psi^{T,L}_{\bar qq}(\vec r,\alpha)=
\frac{i\,Z_q\sqrt{\alpha_{em}}}
{4\pi\,p\,\alpha(1-\alpha)}
\int\limits_{-\infty}^{z_2}dz_1\,
\Bigl(\bar\chi\;\widehat O^{T,L}\chi\Bigr)\,
G^{vac}_{\bar qq}(z_1,\vec r_1;z_2,\vec r_2)
\Bigr|_{r_1=0;\ \vec r_2=\vec r}\ .
\label{2.2}
\eeq
In the limit of vanishing interaction, $a(\alpha)\to 0$
the wave functions (\ref{2.2}) recovers the perturbative
ones Eq.~(\ref{1.4}).

Note that in $k_T$ representation the free Green function
$G^0_{\bar qq}(z_1,\vec r_1;z_2,\vec r_2)$ integrated over longitudinal 
coordinate is simply related to the coherence length (\ref{1.1}), if one
performs an analytic continuation to imaginary time, $z\to -iz$
\beq
\int\limits_{z_1}^{\infty}d\,z_2\,
G^0_{\bar qq}(z_1,\vec r_1;z_2,\vec r_2)=
\,\int \frac{d^2k_T}{(2\,\pi)^2}\ 
{\rm exp}\Bigl[-i\,\vec k_T\cdot(\vec r_2-\vec r_1)\Bigr]
\,l_c(k_T,\alpha)\ .
\label{2.3}
\eeq
This is easily generalized to include the nonperturbative
interaction. Then,
making use of this relation one can switch in (\ref{1.7}) to
$r_T$ representation and express the mean coherence
length via the Green function. Then, we arrive at new
expressions for the functions $N^{T,L}$ and $D^{T,L}$ in
(\ref{1.12}),
\beqn
N^{T,L}&=& \int\limits_0^1 d\alpha
\int d^2r_1\,d^2r_2\,
\left[\Psi_{\bar qq}^{T,L}
\left(\vec r_2,\alpha\right)\right]^*\,
\sigma^N_{\bar qq}(r_2,\alpha)\,
\left(\int\limits_{z_1}^\infty dz_2\, 
G^{vac}_{q\bar q}\left(\vec r_1,z_1;\vec r_2,z_2\right)\right)
\nonumber\\
&\times&\,\Psi_{\bar qq}^{T,L}
\left(\vec r_1,\alpha\right)\,
\sigma^N_{\bar qq}(r_1,\alpha)\ ,
\label{2.4}
\\
D^{T,L}&=& \int_0^1d\alpha\int d^2r\,
\left|\Psi_{q\bar q}^{T,L}\left(\vec r_,\alpha\right)
\sigma^N_{q\bar q}\left(r,s\right)\right|^2\ ,
\label{2.5}\ 
\eeqn
where the nonperturbative $\bar qq$ wave functions are given by (\ref{2.2}).

For a harmonic oscillator potential the Green function is known
analytically,
\beqn
G^{vac}_{q\bar q}\left(\vec r_2,z_1;\vec r_1,z_2\right)
&=&\frac{a^2\left(\alpha\right)}
{2\pi\sinh\left(\omega\,\Delta z\right)}\,
\exp\left[-\frac{\varepsilon^2\,\Delta z}{2\,\nu\,\alpha(1-\alpha)}\right]
\nonumber\\ &\times&\,
\exp\left\{-\frac{a^2\left(\alpha\right)}{2}\left[
\left(r_1^2+r_2^2\right)\coth\left(\omega\,\Delta z\right)
-\frac{2\vec r_1\cdot\vec r_2}{\sinh\left(\omega \,\Delta z\right)}
\right]\right\},
\label{2.6}
\eeqn
where $\Delta z=z_2-z_1$ and
\beq
\omega=\frac{a(\alpha)^2}{\nu\,\alpha(1-\alpha)}\ ,
\label{2.7}
\eeq
is the oscillator frequency.

The light-cone wave functions (\ref{2.2}) modified by the interaction  
can now be calculated explicitly \cite{kst2},
\beqn
\left[\Psi_{q\bar q}^{T}
\left(\varepsilon,\lambda,\vec r_1\right)\right]^*
\Psi_{q\bar q}^{T}\left(\varepsilon,\lambda,\vec r_2\right)
&=&Z_q^2\,\alpha_{em}\left[
m_q^2\,\Phi_{0}^*\left(\varepsilon,\lambda,\vec r_1\right)
\Phi_{0}\left(\varepsilon,\lambda,\vec r_2\right)\right.
\nonumber\\ &+&\left.\Bigl[1-2\alpha\left(1-\alpha\right)\Bigr]\,
\vec\Phi_{1}^*\left(\varepsilon,\lambda,\vec r_1\right)\cdot
\vec\Phi_{1}\left(\varepsilon,\lambda,\vec r_2\right)\right],
\label{2.8}
\eeqn
\beq
\Psi_{q\bar q}^{L}\left(\varepsilon,\lambda,\vec r_2\right)
=Z_q^2\alpha_{em}
4Q^2\alpha^2\left(1-\alpha\right)^2
\Phi_{0}^*\left(\varepsilon,\lambda,\vec r_1\right)
\Phi_{0}\left(\varepsilon,\lambda,\vec r_2\right),
\label{2.9}
\eeq
where the parameter
\beq
\lambda=\frac{2a^2\left(\alpha\right)}{\varepsilon^2}\ ,
\label{2.10}
\eeq
describes the strength of the interaction; and the functions
$\Phi_{0,1}$ read,
\beqn
&& \Phi_{0}^*\left(\varepsilon,\lambda,\vec r_1\right)
\Phi_{0}\left(\varepsilon,\lambda,\vec r_2\right)
=\frac{1}{\left(4\,\pi\right)^2}
\int\limits_0^\infty du\,dt\,\frac{\lambda^2}
{\sinh\left(\lambda t\right)\,\sinh\left(\lambda u\right)}
\nonumber\\
&\times& 
\exp\left[-\frac{\lambda\varepsilon^2r_1^2}{4}\,\coth\left(\lambda t\right)-t
-\frac{\lambda\varepsilon^2r_2^2}{4}\,\coth\left(\lambda u\right)-u\right],
\label{2.11}
\eeqn
\beqn
&& \vec\Phi_{1}^*\left(\varepsilon,\lambda,\vec r_1\right)\cdot
\vec\Phi_{1}\left(\varepsilon,\lambda,\vec r_2\right)
=\frac{1}{\left(2\,\pi\right)^2}\ 
\frac{\vec r_1\cdot\vec r_2}{r_1^2r_2^2}
\int\limits_0^\infty du\,dt
\nonumber\\ &\times& 
\exp\left[-\frac{\lambda\varepsilon^2r_1^2}{4}\,\coth\left(\lambda t\right)-t
-\frac{\lambda\varepsilon^2r_2^2}{4}\,\coth\left(\lambda u\right)-u\right].
\label{2.12}
\eeqn

 It is easy to verify that in the limit of vanishing interaction,
$\lambda\to 0$, the nonperturbative wave-functions reduce to the
perturbative ones. Compared to the expression for $\Phi_{1}$ in \cite{kst2}, we
have integrated by parts over the parameter $u$, or $t$ respectively. This
considerably simplifies 
the expression.

Now we have all ingredients which are necessary to calculate (\ref{1.12}).
Two from the eight remaining integrations, over the angles, can
be performed analytically. We obtain, 
\beqn
N^T&=&m_N\,x_{Bj}
\int_0^1d\alpha\int_0^\infty dr_1\,r_1\,dr_2\,r_2
\int_0^\infty d\Delta z\,
\left[\Psi_{q\bar q}^{T}\left(\varepsilon,\lambda,\vec r_2\right)\right]^*
\Psi_{q\bar q}^{T}\left(\varepsilon,\lambda,\vec r_1\right)
\nonumber\\ &\times&\,
\sigma^N_{q\bar q}\left(r_2,s\right)\,
\sigma^N_{q\bar q}\left(r_1,s\right)\,
\frac{a^2\left(\alpha\right)}
{\sinh\left(\omega\,\Delta z\right)}\ 
\exp\left[-\frac{\varepsilon^2\,\Delta z}
{2\,\nu\,\alpha(1-\alpha)}\right]
\nonumber\\ &\times&\,
{\rm I}_1\left[\frac{a^2\left(\alpha\right)r_1r_2}
{\sinh\left(\omega\,\Delta z\right)}\right]\ 
\exp\left[-\frac{a^2\left(\alpha\right)}{2}
\left(r_1^2+r_2^2\right)
\coth\left(\omega\,\Delta z\right)\right],
\label{2.13}
\eeqn
\beqn
N^L&=&m_N\,x_{Bj}
\int_0^1d\alpha\int_0^\infty dr_1\,r_1\,dr_2\,r_2
\int_0^\infty d\Delta z\,
\left[\Psi_{q\bar q}^{T}\left(\varepsilon,\lambda,\vec r_2\right)\right]^*
\Psi_{q\bar q}^{T}\left(\varepsilon,\lambda,\vec r_1\right)
\nonumber\\ &\times&\,
\sigma^N_{q\bar q}\left(r_2,s\right)\,
\sigma^N_{q\bar q}\left(r_1,s\right)\,
\frac{a^2\left(\alpha\right)}
{\sinh\left(\omega\,\Delta z\right)}\
\exp\left[-\frac{\varepsilon^2\,\Delta z}
{2\,\nu\,\alpha(1-\alpha)}\right]
\nonumber\\ &\times&\,
{\rm I}_0\left[\frac{a^2\left(\alpha\right)\,r_1\,r_2}
{\sinh\left(\omega\,\Delta z\right)}\right]\,
\exp\left[-\frac{a^2\left(\alpha\right)}{2}
\left(r_1^2+r_2^2\right)\coth\left(\omega\,\Delta z\right)
\right]\ ,
\label{2.14}
\eeqn
\beq
D^{L,T}=
\int_0^1d\alpha\int_0^\infty dr\,r\,
\left|
\Psi_{q\bar q}^{T,L}\left(\varepsilon,\lambda,\vec r\right)\,
\sigma^N_{q\bar q}\left(r,s\right)\right|^2.
\label{2.15}
\eeq
For a dipole
cross section that levels off like (\ref{1.9}) at large separations the
integrations over $r_1$ and $r_2$ can also be done analytically.
However, we prefer to work with the more general expressions that hold
for arbitrary $\sigma^N_{q\bar q}\left(s,r\right)$,
as long as it depends only  on the modulus of $r$.
We perform the remaining integrations
numerically. The results for $l_c^{T,L}(x,Q^2)$ are shown by solid
curves in Figs.~\ref{l-Q2} and \ref{l-x}.

\subsection{Coherence length for gluon shadowing}\label{lc-g}

Shadowing in the nuclear gluon distributing function at small 
$x_{Bj}$ which looks like gluon fusion $GG\to G$ in the infinite 
momentum frame of the nucleus, 
should be treated in the rest frame of the nucleus as
shadowing for the Fock components of the photon 
containing gluons. Indeed, the first shadowing term contains
double scattering of the projectile gluon via exchange of two 
$t$-channel gluons, which is the same Feynman graph as
gluon fusion. Besides, both correspond to the triple-Pomeron
term in diffraction which controls shadowing.

The lowest gluonic Fock component is the $|\bar qqG\ra$. 
The coherence length relevant to
shadowing depends according to (\ref{1.1}) on the effective mass
of the $|\bar qqG\ra$ which should be expected to be 
heavier than that for a $|\bar qq\ra$, and even more
for higher Fock components. Correspondingly,
the coherence length $\la l^G_c\ra$ should be
shorter and a onset of gluon shadowing  
is expected to start at smaller $x_{Bj}$.

For this coherence length one can use 
the same Eq.~(\ref{1.1}), but with the effective mass,
\beq
M^2_{\bar qqG}=\frac{k_T^2}{\alpha_G(1-\alpha_G)}+
\frac{M_{\bar qq}^2}{1-\alpha_G}\ ,
\label{3.1}
\eeq
where $\alpha_G$ is the fraction of the photon momentum 
carried by the gluon, and
$M_{\bar qq}$ is the effective mass of the $\bar qq$ pair.
This formula is, however, valid only in the perturbative limit.
It is apparently affected by the nonperturbative interaction of gluons
which was
found in \cite{kst2} to be much stronger than that for a $\bar qq$.
Since this interaction may substantially modify
the effective mass $M_{\bar qqG}$ we switch to
the formalism of Green function described above, which
recovers Eq.~(\ref{3.1}) in the limit of high $Q^2$.

We treat gluons as massless and transverse. For factor $P$ defined in
(\ref{1.1}) one can write,
\beq\label{gluonl}
\Bigl\la P^{G}\Bigr\ra = 
\frac{N^G}{D^G}\ ,
\label{3.2}
\eeq
where 
\beqn\nonumber
N^{G}&=& m_N\,x_{Bj}\,
\int d^2r_{1G}\,d^2r_{1q\bar q}\,d^2r_{2G}\,
d^2r_{2q\bar q}\,d\alpha_q
\,d\ln(\alpha_G)\,
\widetilde \Psi^{\dagger}_{\bar qqG}
\left(\vec r_{2G},\vec r_{2q\bar q},
\alpha_q,\alpha_G\right)\\ & \times &
\left(\int\limits_{z_1}^\infty dz_2\, G_{\bar qqG}
\left(\vec r_{2G},\vec r_{2q\bar q},z_2;
\vec r_{1G},\vec r_{1q\bar q},z_1\right)
\right)\,
\widetilde \Psi_{\bar qqG}\left(\vec r_{1G},\vec r_{1q\bar q},
\alpha_q,\alpha_G\right)
\label{3.3}
\eeqn
\beqn\nonumber
D^{G}&=&\int d^2r_{1G}\,d^2r_{1q\bar q}\,d^2r_{2G}\,
d^2r_{2q\bar q}\,d\alpha_q
\,d\ln(\alpha_G)\,
\widetilde \Psi^{\dagger}_{\bar qqG}
\left(\vec r_{2G},\vec r_{2q\bar q},
\alpha_q,\alpha_G\right)\\ & \times &
\delta^{\left(2\right)}\left(\vec r_{2G}-\vec r_{1G}\right)\,
\delta^{\left(2\right)}
\left(\vec r_{2q\bar q}-\vec r_{1q\bar q}\right)\,
\widetilde \Psi_{\bar qqG}\left(\vec r_{1G},\vec r_{1q\bar q},
\alpha_q,\alpha_G\right)
\label{3.4}
\eeqn
Here we have introduced the Jacobi variables, 
$\vec r_{q\bar q}=\!\vec R_{\bar q}-\vec R_{q}$ and
$\vec r_{G}=\!\vec R_G-(\alpha_{\bar q}\vec R_{\bar q}+\alpha_q\vec R_{q})
/(\alpha_{\bar q}+\alpha_q)$. $\vec R_{G,q,\bar q}$ are the position vectors of
the gluon, the quark and the antiquark in the transverse plane and
$\alpha_{G,q,\bar q}$ are the longitudinal momentum fractions.

The Green function describing propagation
of the $\bar qqG$ system satisfies the time evolution
equation \cite{kst2},
\beqn
&&
\left[\frac{\partial}{\partial z_2}
-\frac{Q^2}{2\nu}
+\frac{\alpha_q+\alpha_{\bar q}}
{2\nu\alpha_q\alpha_{\bar q}}
\Delta_\perp\left(r_{q\bar q}\right)
+\frac{\Delta_\perp\left(r_{2G}\right)}
{2\nu\alpha_G\left(1-\alpha_G\right)}
-V\left(\vec r_{2G},
\vec r_{2q\bar q},\alpha_q,\alpha_G,z_2\right)
\right]
\nonumber\\ &\times&
G_{q\bar qG}
\left(\vec r_{2G},\vec r_{2q\bar q},z_2;
\vec r_{1G},\vec r_{1q\bar q},z_1\right)
\,=\,
\delta\left(z_2-z_1\right)\,
\delta^{\left(2\right)}\left(\vec r_{2G}-\vec r_{1G}\right)\,
\delta^{\left(2\right)}
\left(\vec r_{2q\bar q}-\vec r_{1q\bar q}\right)\ .
\label{3.5}
\eeqn

In order to calculate the coherence
length relevant to shadowing, we employ the amplitude for diffractive
dissociation $\gamma^*\to \bar qqG$, which is 
the  $\bar qqG$ wave function 
weighted by the cross section \cite{kst2},
\beqn\nonumber\lefteqn{
\widetilde \Psi_{\bar qqG}
\left(\vec r_{G},\vec r_{q\bar q},
\alpha_q,\alpha_G\right) =
\Psi^{T,L}_{\bar qq}\left(\vec r_{q\bar q},\alpha_q\right)
\left[
\Psi_{qG}\left(\frac{\alpha_G}{\alpha_q},\vec r_{G}
+\frac{\alpha_{\bar{q}}}{\alpha_q+\alpha_{\bar{q}}}\,
\vec r_{q\bar q}\right)\right.}
&&\\
\nonumber & - & \left.
\Psi_{\bar qG}\left(\frac{\alpha_G}{\alpha_{\bar{q}}},\vec r_{G}
-\frac{\alpha_{q}}{\alpha_q+\alpha_{\bar{q}}}\,
\vec r_{q\bar q}\right)
\right]\ 
\frac{9}{8}\ 
\left[\sigma_{q\bar q}^N\left(x,\vec r_{G}
+\frac{\alpha_{\bar{q}}}{\alpha_q+\alpha_{\bar{q}}}\,
\vec r_{q\bar q}\right)
\right.\\
&&\left.
+\ \sigma_{q\bar q}^N\left(x,\vec r_{G}
-\frac{\alpha_{q}}{\alpha_q+\alpha_{\bar{q}}}\,
\vec r_{q\bar q}\right)
-\sigma_{q\bar q}^N\left(x,r_{q\bar q}\right)
\right]\ .
\label{3.6}
\eeqn

As different from the case of the $|\bar qq\ra$ Fock state,
where perturbative QCD can be safely applied at high $Q^2$,
the nonperturbative effects remain important for the
$|\bar q\,q\,G\ra$ component even for highly virtual photons.
High $Q^2$ squeezes the $\bar qq$ pair down to a size $\sim 1/Q$,
however the mean quark-gluon separation at $\alpha_G \ll 1$ 
depends on the strength of gluon interaction which
is characterized in this limit by the
parameter $b_0\approx 0.65\,GeV$ \cite{kst2}.
For $Q^2\gg b_0^2$ the $\bar qq$ is small,
$r^2_{\bar qq} \ll r_G^2$, and one can treat the $\bar qqG$ system
as a color octet-octet dipole, {\it i.e.},
\beq\label{app1}
G_{q\bar qG}\left(\vec r_{2G},
\vec r_{2q\bar q},z_2;\vec r_{1G},\vec r_{1q\bar q},z_1
\right)\ \Rightarrow\ 
G_{q\bar q}\left(\vec r_{2q\bar q},z_2;
\vec r_{1q\bar q},z_1\right)\,
G_{GG}\left(\vec r_{2G},z_2;\vec r_{1G},z_1\right)\ .
\label{3.7}
\eeq
Such a Green function $G_{GG}$ satisfies the simple 
evolution equation \cite{kst2},
\beqn\label{x}\nonumber{
\left[\frac{\partial}{\partial z_2}-\frac{Q^2}{2\nu}
+\frac{\Delta_\perp\left(r_{2G}\right)}
{2\nu\alpha_G\left(1-\alpha_G\right)}
-\frac{b_0^4\,r_{2G}^2}
{2\nu\alpha_G\left(1-\alpha_G\right)}
\right]
G_{GG}\left(\vec r_{2G},z_2;\vec r_{1G},z_1\right)}\\
\qquad=\delta\left(z_2-z_1\right)
\delta^{\left(2\right)}\left(\vec r_{2G}-\vec r_{1G}\right)
\label{3.8}
\eeqn

Correspondingly, the modified $\bar qqG$ wave function
simplifies too,
\beq\label{app2}
\widetilde \Psi_{\bar qqG}
\left(\vec r_{G},\vec r_{q\bar q},
\alpha_q,\alpha_G\right) \Rightarrow
-\,\Psi^L_{\bar qq}\left(\vec r_{q\bar q},
\alpha_q\right)\ \vec r_{G}\cdot\vec\nabla\,
\Psi_{qG}\left(\vec r_{G}\right)\ 
\sigma_{GG}^N\left(x,r_{G}\right)\ ,
\label{3.9}
\eeq
where the nonperturbative quark-gluon wave function 
has a form \cite{kst2},
\beq
\Psi_{qG}\left(\vec r_{G}\right)=
\lim_{\alpha_G\to 0}\Psi_{qG}\left(\alpha_G,r_{G}\right)=
- \frac{2i}{\pi}\,
\sqrt\frac{\alpha_s}{3}\ 
\frac{\vec e\cdot\vec r_{G}}{r^2_{G}}\,
\exp\left(-\frac{b_0^2}{2}\,r^2_{G}\right)\ ,
\label{3.10}
\eeq
and the color-octet dipole cross section reads,
\beq
\sigma_{GG}^N\left(x,r_{G}\right)
=\frac{9}{4}\,
\sigma_{q\bar q}^N\left(x,r_{G}\right)\ .
\label{3.11}
\eeq

With the approximations given above, the factor $\la P^G\ra$
in (\ref{1.1}) for the gluon coherence length is calculated
in Appendix ~A and has the form,
\beq
\left<P^G\right>=\frac{2}{3\ln(\alpha_G^{max}/\alpha_G^{min})}\ 
\int\limits_{\delta^{min}}^{\delta^{max}}\frac{d\delta}{\delta}
\left[\frac{5}{8\left(1+\delta\right)}
+\frac{7}{8\left(1+3\delta\right)}-\frac{\delta}{\delta^2-1}\left(
\psi\left(2\right)-\psi\left(\frac{3}{2}+\frac{1}{2\delta}\right)\right)
\right],
\label{3.12}
\eeq
where
\beq
\psi\left(x\right)=
\frac{d\ln\Gamma\left(x\right)}{dx}\ ,
\hspace{2cm}
%\label{3.13}
%\eeq
%and,
%\beq
\delta=\frac{2\,b_0^2}{Q^2\,\alpha_G}\ .
\label{3.14}
\eeq

Both the numerator and denominator in (\ref{3.12})
diverge logarithmically for $\alpha_G^{min}\to 0$,
as it is characteristic for radiation of vector bosons. 
To find an appropriate lower cut off, 
note that the mass of the $q\bar qG$ system is
approximately given by
\beq
M^2_{q\bar qG}\approx \frac{2b_0^2}{\alpha_G}+Q^2,
\eeq
where we used eq.~(\ref{3.1})
with $\la k_T^2\ra \approx b_0^2$.
We demand that $M^2_{q\bar qG}<0.2s$ which leads to
$\alpha_G^{min}=2b_0^2/(0.2s-Q^2)$. 
Furthermore
we work in the approximation of $\alpha_G\ll 1$ and we also 
have to choose an upper cut off.
We use 
\beq\label{limitst}
\frac{2b_0^2}{0.2s-Q^2}\le\alpha_G\le\frac{2b_0^2}{Q^2},
\eeq
which means that we take only masses 
$2Q^2\leq M^2_{q\bar qG}\leq 0.2\,s$ into account.
The two limits become equal  at
$x_{Bj}\approx 0.1$.

Our results for $\la P^G\ra = \la l_c^G\ra/l_c^{max}$ 
are depicted in Fig.~\ref{l-Q2}. With approximations made above 
we cannot cover the low $Q^2$ region and perform calculations
at $Q^2>1\,GeV^2$. The found coherence length is much shorter than 
both $l_c^T$ and $l_c^L$ for $|\bar qq\ra$ fluctuations. This
conclusion corresponds to delayed onset of gluon shadowing
shifted to smaller $x_{Bj}$ predicted in \cite{kst2}.

\section{Shadowing for longitudinal and transverse photons}\label{shad}
 
\subsection{\boldmath$\sigma_L/\sigma_T$ on a nucleon target}
\label{lt-nucleon}

As soon as realistic wave functions for $\bar qq$ fluctuations
including the nonperturbative effects are available, as well as
the energy dependent phenomenological dipole cross section,
we are in position to calculate the longitudinal and transverse
cross sections for a proton target covering also the region of
small $Q^2$,
\beq
{\sigma_{T,L}^{\gamma^*p}}=
\int\limits_0^1d\alpha\int d^2r\,
\left|\Psi_{q\bar q}^{T,L}
\left(\vec r_,\alpha\right)\right|^2
\sigma^N_{q\bar q}\left(r,s\right)
\label{4.0}
\eeq

\begin{figure}[tbh]
\includegraphics{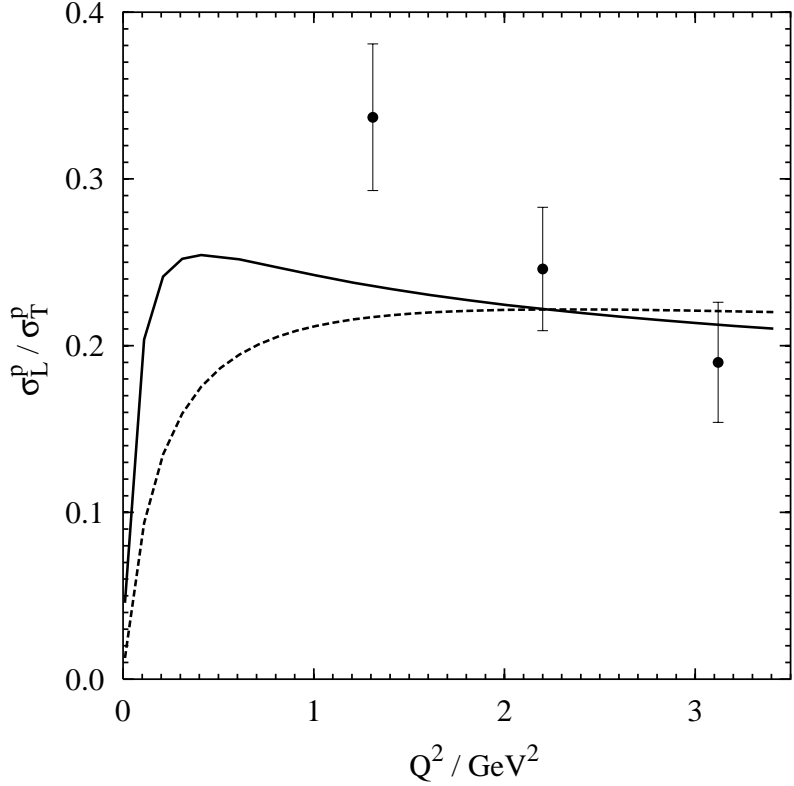}
\begin{center}
\vspace{10.5cm}
\parbox{13cm}
{\caption[shad1]
{\sl Ratio of longitudinal to transverse photoabsorption
cross sections as function of $Q^2$ at $x_{Bj}=0.01$.
The solid curve is calculated with (\ref{4.0}) and the wave functions
including the nonperturbative effects, while for the dashed curve
the perturbative wave functions with $m_q=200\,MeV$ are used.
The NMC data points \cite{nmc1} correspond to $x_{Bj}=0.008,\ 
0.0125,\ 0.0175$ from smaller to higher $Q^2$.}
\label{lt}}
\end{center}
\end{figure}

The results of calculations for the ratio $\sigma_L/\sigma_T$
is shown in Fig.~\ref{lt} by solid curve 
as function of $Q^2$ at $x_{Bj}=0.01$ which is about 
the lowest value of $x_{Bj}$ in the
HERMES data.
As one could expect the ration vanishes towards $Q^2=0$,
however is nearly  constant down to 
very small $Q^2\approx 0.3\,GeV^2$.
To see how well an effective quark mass can mimic
the effect of the nonperturbative interactions 
we have also performed calculations with the perturbative
wave functions and $m_q=200\,MeV$ and plotted the results by
dashed curve in Fig.~\ref{lt}. Comparison demonstrates 
a substantial difference at small $Q^2$.

\subsection{Nuclear targets}\label{lt-a}

Although the coherence length is an important 
characteristics for shadowing, it is not sufficient to
predict nuclear effects in the structure function.
Shadowing for
parton densities at small $x_{Bj}$ in the nuclear structure
function which is defined in the infinite
momentum frame originates from the 
nonlinear effect of parton fusion
in the evolution equation \cite{glr,mq}.
Although the partonic treatment varies from to another 
reference frame, all
observables including shadowing are Lorentz invariant.
In the rest frame of the nucleus shadowing in the total
virtual photoabsorption cross section $\sigma^{\gamma^*A}_{tot}$
(or the structure function $F_2^A$) can be decomposed over 
different Fock components of the photon,
\beq
\sigma_{tot}^{\gamma^*A} = A\,\sigma_{tot}^{\gamma^*N}\,
-\, \Delta\sigma_{tot}(\bar qq)\,
-\, \Delta\sigma_{tot}(\bar qqG)\, 
-\, \Delta\sigma_{tot}(\bar qq2G)\, -\,...
\label{4.1}
\eeq

According to above calculations and Fig.~\ref{lt} the coherence
length corresponding to gluon shadowing is rather small
compared to the mean internucleon spacing in a nucleus
in the kinematical region we are interested in.
Therefore, we hold only the first two terms in the {\it r.h.s.}
of Eq.~(\ref{4.1}), which can be represented for transverse and
longitudinal photons as,
\beqn
&&\Bigl(\sigma^{\gamma^*A}_{tot}\Bigr)^{T,L} = 
A\,\Bigl(\sigma^{\gamma^*N}_{tot}\Bigr)^{T,L}
\,-\, \frac{1}{2} Re\int d^2b
\int\limits_0^1 d\alpha
\int\limits_{-\infty}^{\infty} dz_1 \int\limits_{z_1}^{\infty} dz_2
\int d^2r_1\int d^2r_2\,
\\ & \times &
\Bigl[\Psi^{T,L}_{\bar qq}\left(\varepsilon,
\lambda,r_2\right)\Bigr]^*\,
\rho_A\left(b,z_2\right)\sigma_{q\bar{q}}^N\left(s,r_2\right)
G\left(\vec r_2,z_2\,|\,\vec r_1,z_1\right)\,
\rho_A\left(b,z_1\right)\sigma_{q\bar{q}}^N\left(s,r_1\right)
\Psi^{T,L}_{\bar qq}\left(\varepsilon,\lambda,r_1\right), \nonumber
\label{4.1a}
\eeqn
where $\rho_A(b,z)$ is the nuclear density dependent on impact parameter $b$
and longitudinal coordinate $z$.
The nonperturbative wave functions for the $\bar qq$ component
of the photon are defined in 
Eqs.~(\ref{2.8})-(\ref{2.9}). The Green function
$G\left(\vec r_2,z_2\,|\,\vec r_1,z_1\right)$ describes propagation
of a nonperturbatively  interacting $\bar qq$ pair in an
absorptive medium. It fulfils the evolution equation,
\beqn\nonumber
\left[i\frac{\partial}{\partial z_2}
+\frac{\Delta_\perp\left(r_2\right)-\varepsilon^2}
{2\nu\alpha\left(1-\alpha\right)}
+\frac{i}{2}\rho_A\left(b,z_2\right)\,
\sigma^N_{q\bar{q}}\left(s,r_2\right)
-\frac{a^4\left(\alpha\right)r_2^2}
{2\nu\alpha\left(1-\alpha\right)}
\right]
G\left(\vec r_2,z_2\,|\,\vec r_1,z_1\right)\\
=
i\delta\left(z_2-z_1\right)
\delta^{\left(2\right)}\left(\vec r_2-\vec r_1\right)\ ,
\label{4.2}
\eeqn
where $a(\alpha)$ and $\lambda$ were introduced in Eqs.~(\ref{2.1a}) and 
(\ref{2.10}) respectively. The third term in the {\it l.h.s.} of
(\ref{4.2}) describes absorption of the $\bar qq$ pair 
in the medium of  density 
$\rho_A\left(b,z\right)$ with cross section 
$\sigma^N_{\bar qq}\left(s,r\right)$.

At small $x_{Bj}$ when the coherence length substantially
exceeds, $l_c^{T,L}\gg (z_2-z_1)$ (the nuclear radius) the solution of
Eq.~(\ref{4.2}) very much simplifies, $G(\vec r_2,z_2\,|\,\vec r_1,z_1)
\propto \delta^{\left(2\right)}\left(\vec r_2-\vec r_1\right)$,
{\it i.e.} Lorentz time dilation ``freezes'' variation
of transverse $\bar qq$ separation.

Correspondingly, the total cross
section gets a simple form \cite{zkl,nz},
\beq
\Bigl(\sigma_{tot}^{\gamma^*A}
\Bigr)^{T,L}_{\nu\rightarrow\infty}
= 2\, \int d\alpha\int d^2r
\left|\Psi_{\bar qq}^{T,L}\left(\varepsilon r\right)\right|^2\!
\int d^2b\left[1-\exp\left(-\frac{\sigma^N_{q\bar q}\left(r\right)}{2}
T\left(b\right)\right)\right]\ ,
\label{4.3}
\eeq
where 
\beq
T\left(b\right)=\int\limits_{-\infty}^{\infty} dz\, 
\rho_A\left(b,z\right)
\label{4.4}
\eeq
is the thickness function of the nucleus.

Eq.~(\ref{4.2}) has an explicit analytical solution only if the
dipole cross section $\sigma^N_{\bar qq}(r)= C\,r^2$
and the nuclear density is constant $\rho_A(b,z)=\rho_0$.
Such an approximation has a reasonable accuracy, especially for heavy
nuclei. Nevertheless, it can be even more precise if one makes use 
of the fact that the asymptotic expression (\ref{4.3}) 
is easily calculated with
exact (realistic) dipole cross section and nuclear density.
We need to use the solution of Eq.~(\ref{4.2}) only in the transition
region from no-shadowing to a fully developed shadowing given by
(\ref{4.3}). First of all, we fixed factor $C$ in the simplified 
the dipole cross section demanding to have the same asymptotic 
shadowing (\ref{4.3}) as with realistic one given by (\ref{1.9}).
This was done with the realistic Woods-Saxon form for nuclear density
\cite{ws} and separately for transverse and longitudinal photons and for each
value of $\alpha$.
Then we switched to a constant nuclear density $\rho_0$, 
demanding it to lead 
to the same asymptotic shadowing in (\ref{4.3}) as with the realistic
one. We have checked that the found value of $\rho_0$ is practically
independent of the value of the cross section in the interval $1-50\,mb$.

\begin{figure}[tbh]
\includegraphics{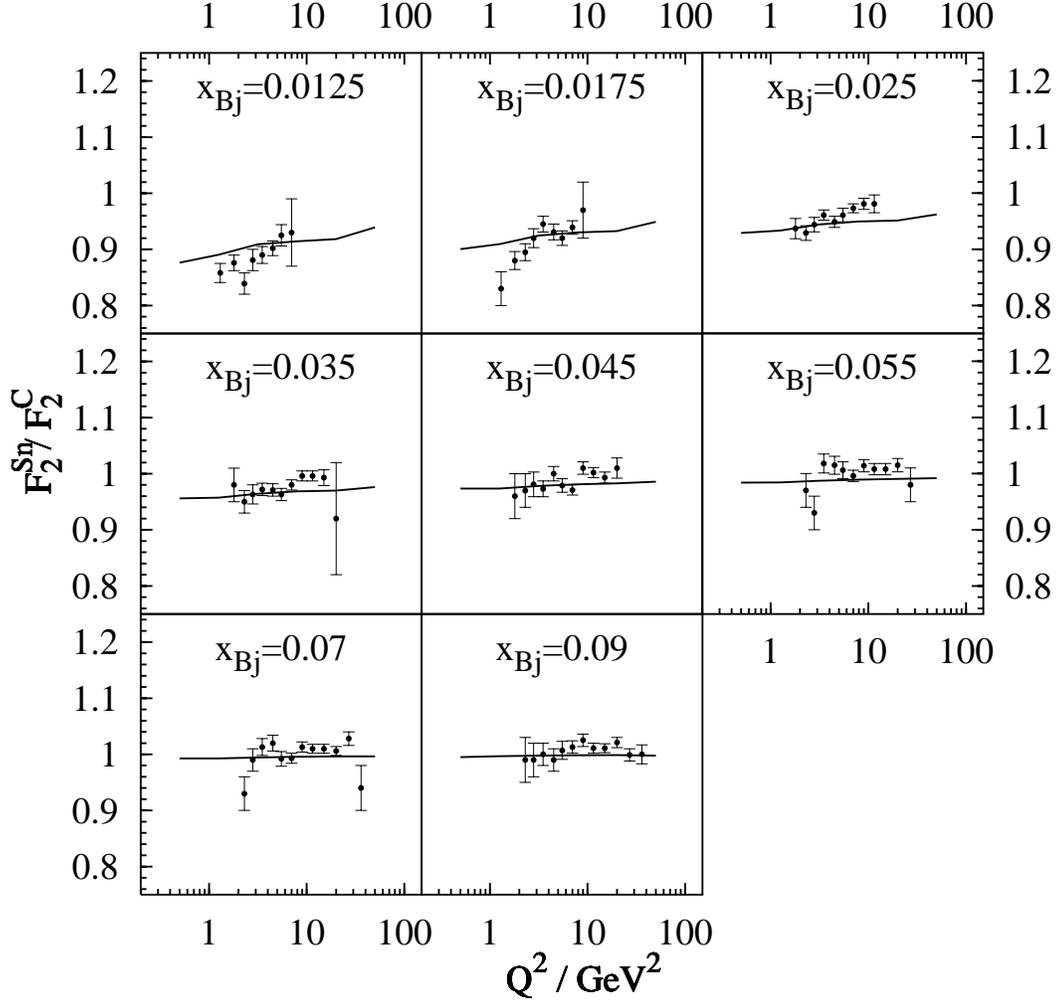}
\begin{center}
\vspace{13.5cm}
\parbox{13cm}
{\caption[shad1]
{\sl $Q^2$ and $x_{Bj}$ dependence of structure function ratio 
for tin to carbon. Full circles show the NMC data \cite{nmc}. 
The crosses show the results of
our calculations for the same kinematics.}
\label{snc}}
\end{center}
\end{figure}

First of all we have checked our formalism comparing with the NMC 
results for shadowing in the nuclear structure function. 
Fig.~\ref{snc} 
demonstrates the data \cite{nmc} for tin to carbon ratio of
proton structure functions
depicted by full circles. 
As was pointed out above gluon shadowing is negligibly small at 
$x_{Bj}>0.01$ which covers the whole range on the NMC experiment.
We performed calculations with parameter $v=0.5$ in (\ref{2.1a}),
but we have checked that the results are independent of $v$.
Although the calculations
are parameter free, agreement is pretty good.

\begin{figure}[tbh]
\includegraphics{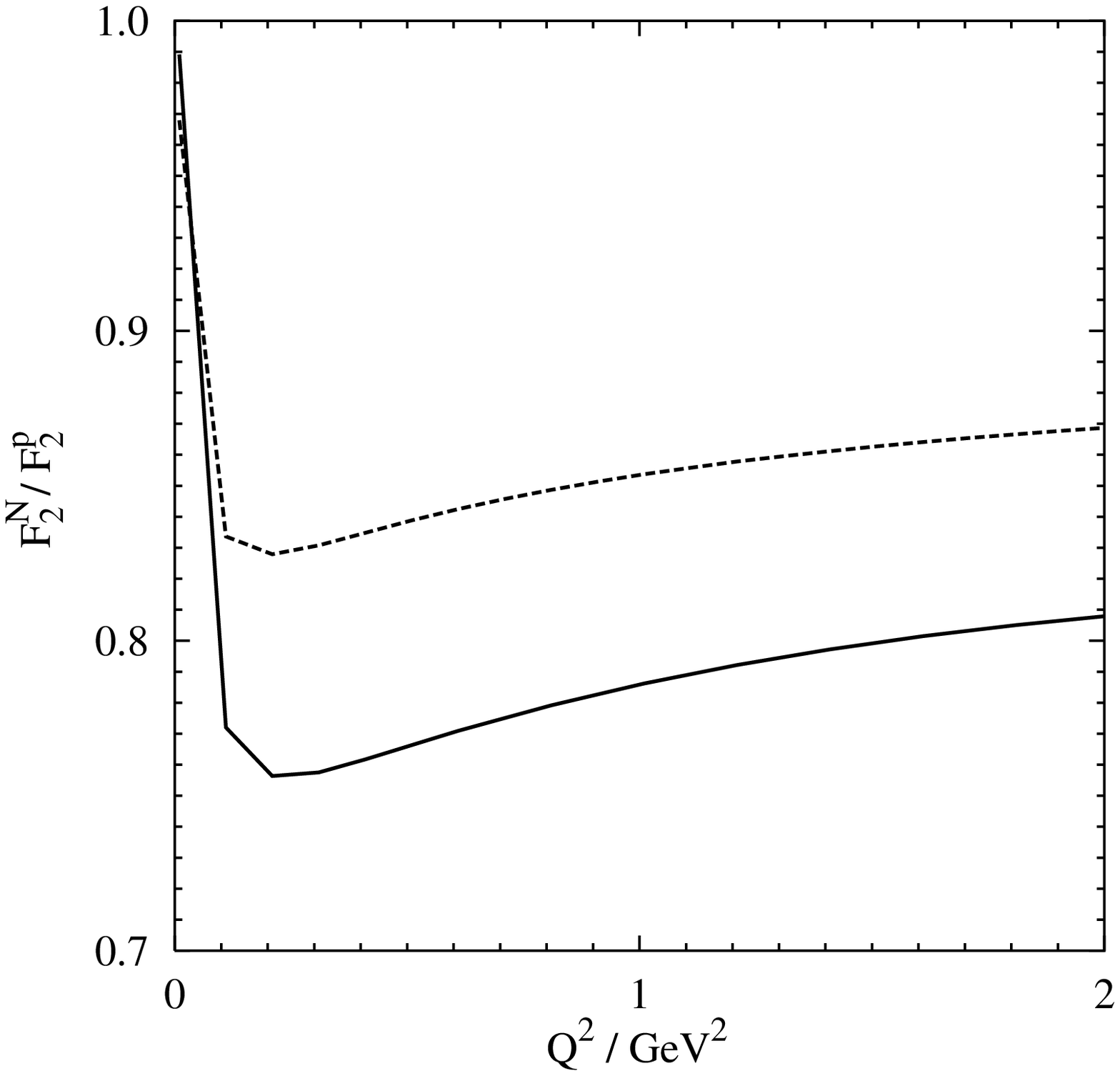}
\begin{center}
\vspace{10.5cm}
\parbox{13cm}
{\caption[shad1]
{\sl The shadowing ratio for nitrogen over proton at low $Q^2$ and
$x_{Bj}=0.01$. Shadowing disappears as $Q^2\to 0$, because of the vanishing
coherence length.}
\label{f2-n}}
\end{center}
\end{figure}

Eventually, we are able to provide predictions for
the kinematical range of HERMES. 
The ratio of the virtual photoabsorption cross sections
for nitrogen to hydrogen at $x_{Bj}=0.01$ versus $Q^2$ is plotted
in Fig.~\ref{f2-n}. 
The solid and dashed curves correspond to the nonperturbative 
and perturbative wave functions respectively.
A salient feature of these predictions is shadowing
vanishing towards $Q^2=0$. This observation does 
not contradict the well known fact of shadowing for real photo absorption, 
but is a rather simple consequence of kinematics. The colliding
energy $s=Q^2\,x_{Bj}$ vanishes along with $Q^2$ at fixed $x_{Bj}$.
This example shows that $x_{Bj}$ is a poor variable at small $Q^2$
and should be replaced by $s$.
This kinematical effect may be partially responsible for the
unusual enhancement of nuclear shadowing with rising $Q^2$ detected
in the HERMES experiment \cite{hermes}. Our calculated nitrogen to proton
ratio of the cross sections reaches its minimum at $Q^2\sim 0.2\,GeV^2$ 
and then smoothly rises with $Q^2$.

\begin{figure}[tbh]
\includegraphics{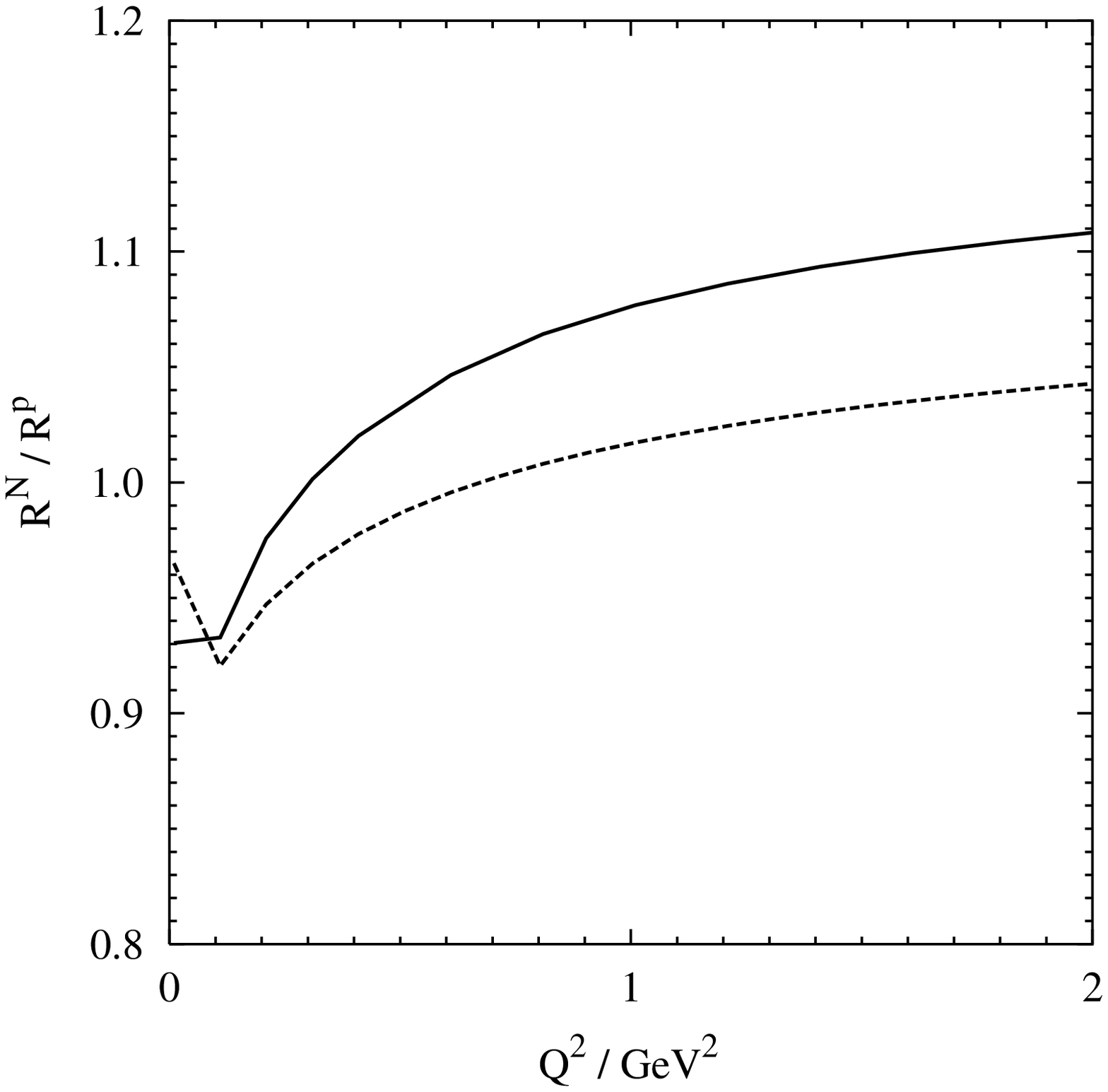}
\begin{center}
\vspace{10.5cm}
\parbox{13cm}
{\caption[shad1]
{\sl Ratio of longitudinal to transverse fractions of the
cross section $R=\sigma_L/\sigma_T$ on nitrogen to proton
targets. The solid and dashed curves correspond to
the photon wave function with and without nonperturbative
effects.}
\label{lt-n}}
\end{center}
\end{figure}

The most striking feature of the HERMES data for shadowing is
a dramatically rising $R=\sigma_L/\sigma_T$ ratio on nitrogen compared to
proton target at $Q^2<1\,GeV^2$ \cite{hermes}. Our predictions 
for $R_N/R_p$ are plotted in Fig.~\ref{lt-n} versus $Q^2$ at
$x_{Bj}=0.01$ where the experimental ratio $R_N/R_p\approx 5$.
Apparently, we do not expect any remarkable effect.
Moreover, this ratio does not change much from proton
to nitrogen in spite of the much longer coherence length
for longitudinal photons predicted above. However,
this effect leading to a stronger shadowing for $\sigma_L$
is compensated by the fact that longitudinal photons
develop fluctuations of a smaller size compared
to that in transverse photons, {\it i.e.} they are less
shadowed. Note that nuclear effects for $R$ were estimated 
previously in \cite{barone} where many of important 
ingredients of present approach were missed.

\section{Summary}

Basing on the light-cone Green function approach 
we provide predictions for nuclear shadowing 
in the most difficult for calculation region
of medium-small $x_{Bj}>0.01$.
Since the nonperturbative effects are included we
are also predict shadowing down to small $Q^2<1\,GeV^2$
which is the kinematical region where the HERMES experiment
discovered unusual shadowing effects. 

We found that the coherence length which controls shadowing
is nearly three time longer for longitudinal than for 
transverse photons, and it is very much different from
what is suggested by the widely accepted approximation.

Using the Green-function light-cone approach including
the nonperturbative effects we calculated nuclear shadowing
for longitudinal and transverse photons. Although the predicted
nuclear shadowing exposes interesting effects in the region
of small $Q^2$, we are unable to explain the dramatic 
phenomena detected in the HERMES experiment.

We suppose that our results 
provide a reliable base line for nuclear effects
in this region. 
The dramatic effects revealed by the HERMES 
experiment probably cannot be explained without involving a new
nonstandard dynamics ({\it e.g.} see in \cite{mbk}).

{\bf Acknowledgements:} 
We are grateful to Andreas Sch\"afer who
initiated this work and
actively participated in it at the early stage, for
numerous and encouraging discussions. 
We thank Antje Br\"ull, Jeroen van Hunen and
Gerard van der Steenhoven for informing us about
the unusual observation in the HERMES experiment
and for many helpful communications and discussions.
We are also thankful to Yuri Ivanov for help with preparing figures.
This work was partially supported by the 
Gesellschaft f\"ur Schwerionenforschung, GSI, grant HD H\"UF T,
by the grant INTAS-97-OPEN-31696 and by the European Network:
Hadronic Physics with Electromagnetic Probes,
Contract No. FMRX-CT96-0008. Most of this work was done
when A.V.T. was employed by Regensburg University.

%%%%%%%%%%%%%%%%%%%%%%%%
\def\appendix{\par
 \setcounter{section}{0}
 \setcounter{subsection}{0}
 \def\thesection{Appendix \Alph{section}}
 \def\thesubsection{\Alph{section}.\arabic{subsection}}
 \def\theequation{\Alph{section}.\arabic{equation}}
 \setcounter{equation}{0}}
%%%%%%%%%%%%%%%%%%%

\appendix

\section{Calculation of the coherence length for the $\boldmath{\bar qqG}$-state}

The calculations are somewhat more cumbersome for the gluon coherence length.
We consider the case $Q^2\gg b_0^2$. With the approximations (\ref{app1})
to (\ref{app2}) we obtain for the denominator of (\ref{gluonl})
\beq
D^G=\int d^2r_{G}d^2r_{q\bar q}\int_0^1 d\alpha_q 
\int\limits_{\alpha_G^{min}}^{\alpha_G^{max}}\frac{d\alpha_G}{\alpha_G}
\left|\Psi^L\left(r_{q\bar q}\alpha_{q\bar q}\right)\right|^2
\left[\sigma^N_{q\bar q}\left(s,r_G\right)\right]^2
\left[\vec r_{q\bar q}\cdot\vec\nabla\Psi_{qG}\left(r_{G}\right)\right]^2.
\eeq
This integral diverges logarithmically for $\alpha_G^{min}\to 0$. 
To find an appropriate lower cut off, 
note that the mass of the $q\bar qG$ system is
approximately given by
\beq
M^2_{q\bar qG}\approx \frac{2b_0^2}{\alpha_G}+Q^2.
\eeq
We demand that $M^2_{q\bar qG}<0.2s$ which leads to
$\alpha_G^{min}=2b_0^2/(0.2s-Q^2)$. 
Furthermore
we work in the approximation of $\alpha_G\ll 1$ and we also 
have to choose an upper cut off.
We use 
\beq\label{limits}
\frac{2b_0^2}{0.2s-Q^2}\le\alpha_G\le\frac{2b_0^2}{Q^2}
\eeq
The two limits become equal  at
$x_{Bj}\approx 0.1$.
In our further calculation of $D^G$ we do the replacement 
$r_{q\bar qi}r_{q\bar qj}\to r_{q\bar q}^2\delta_{ij}$ and perform the
derivative. This yields
\beqn\nonumber
D^G&=&\left(\frac{2}{\pi}\sqrt{\frac{\alpha_s}{3}}\right)^2
\frac{6\alpha_{em}}{\left(2\pi\right)^2}4Q^2\pi
\int d^2r_{G}dr_{q\bar q}r^3_{q\bar q}\int_0^1 d\alpha_q 
\int\limits_{\alpha_G^{min}}^{\alpha_G^{max}}\frac{d\alpha_G}{\alpha_G}
K_0^2\left(\varepsilon r_{q\bar q}\right)\alpha_q^2\left(1-\alpha_q\right)^2
\\
&\times&\left[\sigma^N_{q\bar q}\left(s,r_G\right)\right]^2
e^{-b_0^2r^2_{G}}\left(\frac{2}{r^4_{G}}+
\frac{2b_0^2}{r^2_{G}}+b_0^4\right).
\eeqn
For the integration over $r_{q\bar q}$ we use the integral representation
\beq
K_0(x)=\frac{1}{2}\int_0^\infty\frac{dt}{t}\exp\left(-t-\frac{x^2}{4t}\right)
\eeq
of the modified Bessel function and obtain
\beq\label{integral}
\int_0^\infty dr_{q\bar q}r^3_{q\bar q}
K_0^2\left(\varepsilon r_{q\bar q}\right)
=\frac{2}{3\varepsilon^4}.
\eeq
Thus we have for the denominator
\beq
D^G=\frac{32\alpha_{em}\alpha_s}{3\pi^2Q^2}
\ln\frac{\alpha_G^{max}}{\alpha_G^{min}}
\int_0^\infty dr_Gr_G
\left[\sigma^N_{q\bar q}\left(s,r_G\right)\right]^2
e^{-b_0^2r^2_{G}}\left(\frac{2}{r^4_{G}}+
\frac{2b_0^2}{r^2_{G}}+b_0^4\right).
\eeq
Now we restrict ourselves to the a dipole cross section of the form
\beq
\sigma^N_{q\bar q}\left(s,r_G\right)=C(s)r_G^2
\eeq
and perform the last integration with the result
\beq
D^G=\frac{32\alpha_{em}\alpha_sC^2(s)}{\pi^2Q^2b_0^2}
\ln\frac{\alpha_G^{max}}{\alpha_G^{min}}.
\eeq
Note that the factor $C(s)$ will drop out, when one takes the ratio 
$<P^G>=N^G/D^G$.

Next we  calculate the numerator
\beqn\nonumber
N^G&=&m_Nx_{Bj}\left(\frac{2}{\pi}\sqrt{\frac{\alpha_s}{3}}\right)^2
\frac{6\alpha_{em}}{\left(2\pi\right)^2}4Q^2
\int d^2r_{1G}d^2r_{2G}d^2r_{q\bar q}\int_0^1d\alpha_q 
\int\limits_{\alpha_G^{min}}^{\alpha_G^{max}}\frac{d\alpha_G}{\alpha_G}
\alpha_q^2\left(1-\alpha_q\right)^2\\
\nonumber&\times&K_0^2\left(\varepsilon r_{q\bar q}\right)
\sigma^N_{q\bar q}\left(s,r_{1G}\right)\sigma^N_{q\bar q}\left(s,r_{2G}\right)
\left[\vec r_{q\bar q}\cdot\vec\nabla_{r_{1G}}
\frac{\vec e\cdot\vec r_{1G}}{r^2_{1G}}e^{-\frac{b_0^2r^2_{1G}}{2}}\right]
\left[\vec r_{q\bar q}\cdot\vec\nabla_{r_{2G}}
\frac{\vec e\cdot\vec r_{2G}}{r^2_{2G}}e^{-\frac{b_0^2r^2_{2G}}{2}}\right]\\
&\times&
\int_0^\infty d\Delta z G_{GG}\left(\vec r_{2G},\vec r_{1G},\Delta z\right)
\eeqn
where
\beq
G_{GG}\left(\vec r_{2G},\vec r_{1G},\Delta z\right)
=\frac{b_0^2e^{-\frac{Q^2\Delta z}{2\nu}}}
{2\pi\sinh\left(\omega\Delta z\right)}
\exp\left\{-\frac{b_0^2}{2}\left[\left(r^2_{1G}+r^2_{2G}\right)
{\rm cth}\left(\omega\Delta z\right)
-\frac{2\vec r_{1G}\cdot\vec r_{2G}}{\sinh\left(\omega\Delta z\right)}
\right]\right\}
\eeq
is the solution of (\ref{x}) with $\omega=b_0^2/(\nu\alpha_G)$.

Again we can do the replacement
$r_{q\bar qi}r_{q\bar qj}\to r_{q\bar q}^2\delta_{ij}$, perform the
derivatives, sum over gluon polarizations and use (\ref{integral}) to obtain
\beqn\nonumber
N^G&=&m_Nx_{Bj}\frac{8\alpha_{em}\alpha_s}{3\pi^4Q^2}
\int d^2r_{1G}d^2r_{2G}d\Delta z
\int\limits_{\alpha_G^{min}}^{\alpha_G^{max}}\frac{d\alpha_G}{\alpha_G}
\sigma^N_{q\bar q}\left(s,r_{1G}\right)\sigma^N_{q\bar q}\left(s,r_{2G}\right)
\frac{b_0^2e^{-\frac{Q^2\Delta z}{2\nu}}}
{\sinh\left(\omega\Delta z\right)}
\\&\times&\nonumber
\left[
4\frac{\left(\vec r_{1G}\cdot \vec r_{2G}\right)^2}{r^4_{1G}r^4_{2G}}
+b_0^4\frac{\left(\vec r_{1G}\cdot \vec r_{2G}\right)^2}{r^2_{1G}r^2_{2G}}
+2b_0^2\frac{\left(\vec r_{1G}\cdot \vec r_{2G}\right)^2}{r^2_{1G}r^4_{2G}}
+2b_0^2\frac{\left(\vec r_{1G}\cdot \vec r_{2G}\right)^2}{r^4_{1G}r^2_{2G}}
\right.\\&&\left.
-\frac{b_0^2}{r^2_{1G}}-\frac{b_0^2}{r^2_{1G}}-\frac{1}{r^2_{1G}r^2_{2G}}
\right]
\exp\left(-\beta\left(r^2_{1G}+r^2_{2G}\right)
+2\gamma\vec r_{1G}\cdot \vec r_{2G}\right)
\eeqn
where
\beq
\beta=\frac{b_0^2}{2}\left(1+\coth\left(\omega\Delta z\right)\right),
\eeq
\beq
\gamma=\frac{b_0^2}
{2\sinh\left(\omega\Delta z\right)}.
\eeq
With a cross section like $\sigma^N_{q\bar q}\left(s,r\right)=C(s)r^2$ 
the integrations over $r_{G}$ are easily
performed with the result
\beqn\nonumber
N^G&=&m_Nx_{Bj}\frac{8\alpha_{em}\alpha_sC^2(s)}{3\pi^2Q^2b_0^2}
\int d\Delta z
\int\limits_{\alpha_G^{min}}^{\alpha_G^{max}}\frac{d\alpha_G}{\alpha_G}
\frac{e^{-\frac{Q^2\Delta z}{2\nu}}}
{\sinh\left(\omega\Delta z\right)}
\left\{\frac{10}{\left(1+\coth\left(\omega\Delta z\right)\right)^2}
\right.\\
&&\left.
+\frac{12}{\sinh^2\left(\omega\Delta z\right)
\left(1+{\rm cth}\left(\omega\Delta z\right)\right)^3}
-8\sinh^2\left(\omega\Delta z\right)
\ln\left(\frac{1+\coth\left(\omega\Delta z\right)}{2}\right)
\right\}\\
\nonumber
&=&\frac{8\alpha_{em}\alpha_sC^2(s)}{3\pi^2Q^2b_0^2}
\int_0^1 dy 
\int\limits_{\delta^{min}}^{\delta^{max}}\frac{d\delta}{\delta^2}
\left\{4y^{\frac{1}{\delta}-2}\left(1-y^2\right)\ln\left(1-y^2\right)
+5y^{\frac{1}{\delta}}\left(1-y^2\right)
+12y^{\frac{1}{\delta}+2}
\right\}.\\
\eeqn
In the last step we have introduced the new variables $y=e^{-\omega\Delta z}$
and $\delta=2b_0^2/(Q^2\alpha_G)$. 
According to (\ref{limits}) the limits for $\delta$ are
\beq
1\le\delta\le 0.2\frac{s}{Q^2}-1.
\eeq
For the integral containing the logarithm it
is convenient to do one more substitution, $x=y^2$. Then one finds
\beqn
\int_0^1 dy 
y^{\frac{1}{\delta}-2}\left(1-y^2\right)\ln\left(1-y^2\right)
&=&
\frac{1}{2}\lim_{\eta\to 0}\frac{\partial}{\partial\eta}
\int_0^1 dx x^{\frac{1}{2\delta}-\frac{3}{2}}
\left(1-x\right)^{1+\eta}\\
&=&\frac{2\delta^2}{\delta^2-1}\left(
\psi\left(\frac{3}{2}+\frac{1}{2\delta}\right)
-\psi\left(2\right)\right).
\eeqn
Now only one integration is left in $N^G$
\beq
N^G=\frac{64\alpha_{em}\alpha_sC^2(s)}{3\pi^2Q^2b_0^2}
\int\limits_{\delta^{min}}^{\delta^{max}}\frac{d\delta}{\delta}
\left[\frac{5}{8\left(1+\delta\right)}
+\frac{7}{8\left(1+3\delta\right)}-\frac{\delta}{\delta^2-1}\left(
\psi\left(2\right)-\psi\left(\frac{3}{2}+\frac{1}{2\delta}\right)\right)
\right]
\eeq
and we end up with the result (\ref{3.12}) for the factor
$\la P^G\ra=N^G/D^G$ at $Q^2\gg b_0^2$.


\begin{thebibliography}{99}

\bibitem{hermes}  K. Ackerstaff et al. (The HERMES Collaboration),
{\sl Nuclear Effects on $R = \sigma_L / \sigma_T$ in Deep-Inelastic
Scattering}, hep-ex/9910071.

\bibitem{glr} L.V.~Gribov, E.M.~Levin and M.G~Ryskin, Nucl. Phys. {\bf B188}
(1981) 555; Phys. Rep. {\bf 100} (1983) 1.

\bibitem{mq} A.H.~Mueller and J.W.~Qiu, Nucl. Phys. {\bf B268} (1986) 427.

\bibitem{bauer}
T.H.~Bauer,
R.D.~Spital,
D.R.~Yennie
and
F.M.~Pipkin,
Rev.~Mod.~Phys.~{\bf
50}
(1978)
261.

\bibitem{fs} L.L.~Frankfurt and M.I.~Strikman,
Phys. Rept.{\bf 160} (1988) 235.

\bibitem{kst2}  B.Z.~Kopeliovich, A.~Sch\"afer and A.B.~Tarasov,
  {\sl Nonperturbative Effects in Gluon Radiation
  and Photoproduction of Quark Pairs}, hep-ph/9908245.


\bibitem{zkl} Al.~B.~Zamolodchikov, B.Z.~Kopeliovich and L.I.~Lapidus,
Sov.~Phys.~JETP Lett.~{\bf 33} (1981) 612.

\bibitem{bm} S.J.~Brodsky and A.~Mueller,
Phys. Lett. {\bf
B206}
(1988)
685.

\bibitem{bbgg} G. Bertsch, S.J. Brodsky, A.S. Goldhaber
and
J.F. Gunion,
Phys. Rev. Lett. {\bf 47},
297.

\bibitem{ajm} J.\ M.~Bjorken, J.\ B.~Kogut, 
{\bf D8} (1973) 1341.

\bibitem{krt} B.Z.~Kopeliovich, J.~Raufeisen and A.V.~Tarasov,
 Phys. Lett. {\bf B440} (1998) 151.

\bibitem{rtv} J.~Raufeisen, A.V.~Tarasov and O.~Voskresenskaya,
Eur. Phys. J. {\bf A5} (1999) 173.

\bibitem{zakh} B.G.~Zakharov, Phys. Atom. Nucl. {\bf 61} (1998) 838

\bibitem{ks} J.B.~Kogut and D.E.~Soper,
Phys. Rev. {\bf D1} (1970) 2901.

\bibitem{bks} J.M.~Bjorken, J.B.~Kogut and D.E.~Soper, {\bf D3} (1971) 1382.

\bibitem{nz} N.N.~Nikolaev and B.G.~Zakharov,
Z. Phys. {\bf C49} (1991) 607.

\bibitem{gw} K. Golec-Biernat and M. W\"usthoff,
Phys.\ Rev.\ {\bf D59} (1999) 014017,
Phys.\ Rev.\ {\bf D60} (1999) 114023.

\bibitem{gr} I.S.~Gradshteyn and I.M.~Ryzhik, {\sl Table of Integrals,
Series, and Products},
Academic Press, Inc., Harcourt \& Company, Publishers.

\bibitem{nmc1} The NMC Coll., M.~Arneodo et al.,
Nucl. Phys. {\bf B483} (1997) 3

\bibitem{ws} H.~De~Vries, C.W.~De Jager and C.~De~Vries,
 Atomic Data and Nucl. Data Tables, {\bf 36} (1987) 469.

\bibitem{nmc} The NMC Coll., M.\ Arneodo et al.\, Nucl. Phys.\ {\bf B481} (1996) 23.

\bibitem{barone} V.~Barone and M.~Genovese,  
{\sl  Longitudinal and Transverse Nuclear Shadowing}, hep-ph/9610206;
V.~Barone et al., Phys. Lett. {\bf B304} (1993) 176

\bibitem{mbk}  G.~A.~Miller, S.~J.~Brodsky and M.~Karliner,
{\sl  Coherent Contributions of Nuclear Mesons to Electroproduction
and the HERMES Effect}, hep-ph/0002156.

\end{thebibliography}
\end{document}